# GAMMA IMAGERS FOR NUCLEAR SECURITY AND NUCLEAR FORENSICS
*Recommendations based on results from a side-by-side intercomparison*

L.E. SINCLAIR[1,3], P.R.B. SAULL[2,3], A. MCCANN[1], A.M.L. MACLEOD[2], N.J. MURTHA[3], A. EL-JABY[4] and G. JONKMANS[5]

[1]Canadian Hazards Information Service, Natural Resources Canada, Ottawa, Canada
Email: laurel.sinclair@canada.ca

[2]Metrology Research Centre, National Research Council Canada, Ottawa, Ontario, Canada

[3]Physics Department, Carleton University, Ottawa, Canada

[4]Directorate of Security and Safeguards, Canadian Nuclear Safety Commission, Ottawa, Canada

[5]Centre for Security Science, Defence Research and Development Canada, Ottawa, Canada

**Abstract**

Nuclear security operations and forensic investigations require the utilization of a suite of instruments ranging from passive gamma spectrometers to high-precision laboratory sample analyzers. Gamma spectroscopy survey is further broken down into wide-area search performed with large-volume scintillator-based mobile survey spectrometers which are integrated with geographic position sensors for mapping and identification of hot zones, and high-precision long-dwell measurements using solid state spectrometers for follow-on characterization to establish isotopic content and ratios. While performing well at detecting the presence, quantity and type of radioactivity, all of these methods have limited ability to determine the location of a source of radioactivity. In recent years, technology advances have resulted in gamma imager devices which can create an image of the distribution of radioactive sources using the gamma emissions which accompany radioactive decay, and overlay this on an optical photograph of the environment. These gamma imaging devices have arisen out of methods developed for medical physics, experimental particle physics, and astrophysics, resulting in a proliferation of different technological approaches. Those responsible for establishing a nuclear security concept of operations, require guidance to choose the proper gamma imager for each of the application spaces in a tiered response. Here the results of an intercomparison of two gamma imagers based on two widely different technologies, semiconductor and scintillator detectors, are presented. The optimal utilization of these imaging technologies in a tiered response is discussed based on the results of the trial. Finally, an outlook on future directions for gamma imaging advances is provided.

1. INTRODUCTION

A variety of detection instruments may be employed in a nuclear security or forensic operation to detect and locate nuclear and other radioactive material which is outside of regulatory control [1]. Owing to their ubiquity and penetrating ability, the detection of gamma rays is the cornerstone of methods to locate radioactive material which may be hidden from view some distance away. In recent years there have been significant advances in ability to characterize materials from their gamma emissions by improvement of existing functionality, such as development of materials with improved energy resolution, and reduction of the dead time caused by electronics to near zero even in multi-channel systems, see for example in [2],[3]. The advances with the greatest potential to positively impact nuclear security operations and forensic investigations, however, introduce new functionality: the ability to point in the direction of a radioactive source, and the ability to image or map a radioactivity distribution from a fixed position.

Gamma imager development is currently proceeding rapidly with several competing technologies showing promise. Early gamma imaging designs utilized a heavy inert mask to collimate the gamma rays and improvements to this approach are still taking place [4],[5]. Later designs have been based on Compton imaging, which permits an unlimited (4π) field of view and utilizes electronic collimation, negating the need for transportation of dead material in the form of a mask. There are Compton imagers based on gaseous [6] and noble liquid [7] time-projection chambers, semiconductor detectors [8],[9], and scintillator detectors [10],[11]. All of these devices perform the exciting and impactful new function of being able to show an image of the radioactivity, overlaid on a photograph of the surroundings. However, there has not been standardization of reporting of their performance measures. It can be difficult therefore for an operator to understand which technology best suits his or her application. In fact, in scenarios both prior to and after release of radioactivity, a tiered approach is commonly taken in which a wide-area search performed by high-sensitivity instruments mounted on aerial or ground-based vehicles to locate areas of anomalous radioactivity would be followed by characterization surveys which may include long-dwell high-resolution in-situ measurements, or sampling and laboratory analysis [12]-[14]. It is reasonable to expect that gamma imaging will make strong improvements in the quality of information obtained in all of these operational environments with different technologies being ideally suited for each of them. To assist the operators with choosing the optimal gamma imager for each of the mission spaces, quantitative intercomparison is needed.

The performance of two imagers based on different technology was compared in [15]. However, in that comparison the two devices each collected data at their home laboratories under different background conditions. Here, we present a preliminary intercomparison of the same two imagers, this time using data collected with the two devices at the same location in the identical experimental configuration. We find that the two imagers are each best suited to different tiers of a nuclear security or forensic operation. The imager based on scintillator technology is well suited to wide-area search and the imager based on a solid-state detector is best suited to follow-on in-situ characterization.

## 2. METHOD

### 2.1. Instruments

Two instruments have taken part in the preliminary intercomparison presented here, the H3D H420 [16] and the SCoTSS 3×3 [15]. We note that the SCoTSS technology has been designed and developed by some of the authors of this manuscript. Fig. 1 shows the experimental setup during data taking. The H3D H420 gamma imager is the smaller blue and silver instrument mounted in the detector experimental position on the wooden turntable which is on the cart under the blue canopy in the centre of the photo. The SCoTSS 3×3 instrument, not partaking in experimental measurements in this photo, is the much larger black instrument with a white lid, sitting on a table under the red and white canopy on the right.

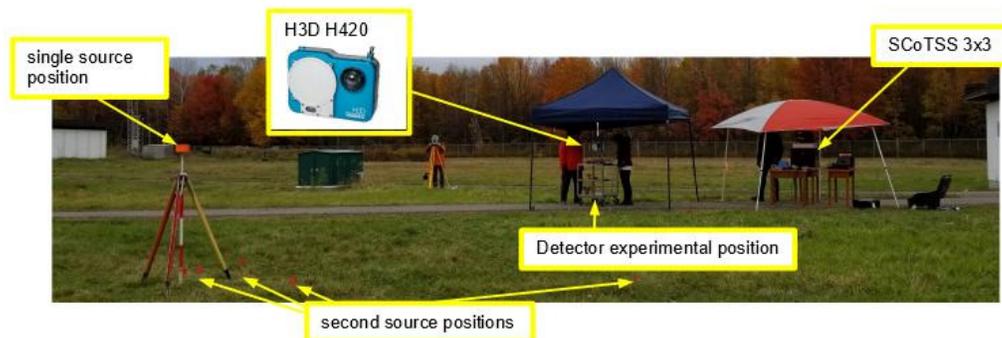

*FIG. 1 Experimental setup. One source is held on the white and red range pole which is visible to the left, held in place by a wooden red and yellow tripod. The red flag markers of some of the surveyed positions for the second source in two-source trials are visible. The detector experimental position for measurements is on the wooden turntable on the metal cart under the blue canopy in the centre of this picture. The H3D H420 detector is the small grey and blue device in the experimental position in the centre of this picture. The SCoTSS 3×3 detector is sitting on a wooden table to the right of the experimental position, under the red and white canopy, not taking part in measurements in this image. A Nikon Nivo Total Station which was used to confirm distance measurements can be seen on the yellow tripod in the distance.*

#### 2.1.1. H3D H420

The H3D H420 imager has been described elsewhere [16],[17]. We repeat here the salient points necessary for understanding of this work. The H3D H420 is a commercially available imager based around a > 19 mm$^3$ cadmium zinc telluride semi-conductor detector. The manufacturer's specifications state that it provides an energy resolution of ≤ 1.1% full-width at half-maximum (FWHM) at 662 keV and that its angular resolution is between 20º and 30º depending on processing method, over the entire 4π field of view. It has a total mass of 3.5 kg and its largest exterior dimension is 24 cm.

*2.1.2. SCoTSS 3×3*

The SCoTSS 3×3 imager has been described in [15]. It is based on 288 individual crystals of CsI(Tl) spectrometer in a "scatter-layer" 12×12 array of 1.35 cm per side cubic crystals and an "absorber layer" 12×12 array of 2.8 cm per side cubic crystals. Each crystal is independently read out with silicon photomultipliers. The SCoTSS 3×3 imager has similar stopping power to the typical 4 L "log" NaI(Tl) crystal used in mobile gamma spectrometry survey as in for example [18],[19]. Energy resolution with the SCoTSS 3×3 imager reaches 7% FWHM at 662 keV, somewhat better than a typical mobile survey spectrometer. The field of view of the SCoTSS 3×3 imager is also unlimited, i.e. 4π. With the SCoTSS 3×3 two-plane design, angular resolution is better than 4% for Cs-137 sources in forward locations, worsening to about 20% to 30% for sources to the side of the imager.

**2.2. Experimental setup**

The SCoTSS 3×3 and H3D H420 imagers were deployed in an intercomparison trial from Oct 16 to 18, 2019, in Ottawa, Canada. One common experimental position (indicated in Fig. 1) was chosen for the detectors. A Hemisphere S321 Global Navigation Satellite System (GNSS) smart antenna was used to obtain the geographic coordinates of this detector experimental position. A source geographic location 10 m from the detector experimental position was then calculated and this position (labelled single source position in Fig. 1) was found using a second Hemisphere S321 GNSS antenna position. A number of geographic locations for the positions of a second source were then calculated at a variety of azimuthal angles also at 10 m from the detector position, and those locations were found using the GNSS antennae. Some of the red indicator flags of these second source positions are apparent in Fig. 1.

The two detectors were mounted alternately on the detector experimental position, on a rotating plate, with material utilized to assure that the centres of their sensitive masses were at the same height. Laser levels were used to extend the horizontal plane from the centre of mass of the sensitive volumes of the imagers to the source locations. The sources were deployed on range poles with the height of the range pole adjusted at each source location to bring the source to the same height as the centres of the sensitive volumes allowing for topographic variability.

A Nikon Nivo Total Station, visible in Fig. 1, was used to verify some of the positions and angles. We estimate that the process followed to set up the trial resulted in an uncertainty of about 5 cm on positions and about half a degree on angles.

**2.3. Datasets**

The runs which supplied data to be discussed herein are listed in Table. 1. Two stainless steel-encapsulated point-like Cs-137 sources each with approximate activity of 160 MBq were utilized. Runs 1 to 4 are single-source runs. Data was collected with the source fixed at the single source position indicated in Fig. 1 and the detectors rotated such that the source appeared at -20º and -120º from their symmetry axes. Runs 6 through 10 are two-source runs. The detectors were rotated such that one source position was at -20º with respect to the detector symmetry axes. Then a second source was placed at -30º, -50º and -120º degrees off axis. Note that the preliminary results discussed herein come from only a subset of the data collected in the Oct. 2019 intercomparison trial and the data were not collected in the order they appear in Table 1.

TABLE 1.    EXPERIMENTAL DATASETS

| Run number | Detector | first source azimuthal angle | second source azimuthal angle |
|---|---|---|---|
| 1 | H3D H420 | -20º | n/a |
| 2 | SCoTSS 3×3 | -20º | n/a |
| 3 | H3D H420 | -120º | n/a |
| 4 | SCoTSS 3×3 | -120º | n/a |
| 5 | H3D H420 | -20º | -30º |
| 6 | SCoTSS 3×3 | -20º | -30º |
| 7 | H3D H420 | -20º | -50º |
| 8 | SCoTSS 3×3 | -20º | -50º |
| 9 | H3D H420 | -20º | -120º |
| 10 | SCoTSS 3×3 | -20º | -120º |

## 2.4. Real-time images

Examples of the results displayed during the data collection are shown in Fig. 2. These are images which were produced by the instruments during the single-source run with the detectors rotated such that the source lay at -20º off-axis. Both detectors successfully localize the source and assign it to a direction just to the left of the canopy pole and consistent with the tripod and range pole location in the approximately two-minute acquisition period shown here. However, it is very difficult to compare the quality of these images. The two detectors use different optical cameras with the H3D H420 optical image having a wider field of view than that of the camera employed by the SCoTSS 3×3 imager. The H3D H420 detector displayed the gamma image in a window which appears to encompass all 4π while the SCoTSS 3×3 display includes only the field of view of the optical camera. The two systems are not consistent in their image processing methods, nor in the levels chosen for the colour contours. In the following subsections we describe the processing method chosen to eliminate these potential biases and present a fair comparison of the two instrument responses.

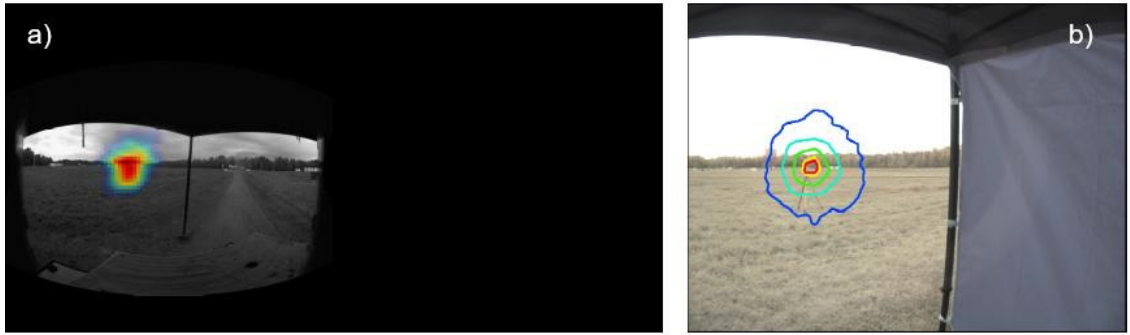

*FIG. 2. Gamma images as displayed during Runs 1 and 2 of the field trial for one 160 MBq Cs-137 source 10 m distant and -20º off axis and approximately two minutes data acquisition. a) The image displayed by the H3D H420 instrument during Run 1. b) The image displayed by the SCoTSS 3×3 instrument during Run 2.*

## 2.5. Event selection and sequencing

Following each run, the H3D H420 and SCoTSS 3×3 instruments both output a data file containing the locations and values of each energy deposition in the detector - a type of data file commonly referred to as "list-mode". The H3D H420 list-mode data and the SCoTSS 3×3 list-mode data were processed through the same analysis code, using the same event selection to the extent possible:
  (a) No lower energy threshold on individual energy deposits was applied by the analysis other than that used in run-time by the instrument.
  (b) Exactly two energy deposits must be recorded in the event.
  (c) The summed energy from the two energy deposits was required to lie within about three sigma of the Cs-137 662 keV photopeak.
     (i)  For H3D H420 the photopeak energy window was 647 keV to 677 keV;

   (ii) For SCoTSS 3×3 the photopeak energy window was 603 keV to 721 keV.
(d) An additional "lever arm" selection was applied for the SCoTSS 3×3 imager. The location of an energy deposit in this detector is assigned to the centre of the 1.35 cm per side or 2.8 cm per side cubic crystal in which it lies. In order to minimize the effect on reconstructed angles of the smearing of the energy deposit position associated with crystal dimension, the two crystal centres are required to lie at least 10 cm apart. No equivalent selection could be found which would improve the H3D H420 images and discussion with the manufacturer indicated that H3D has not found it beneficial in image reconstruction with these detectors to make use of such a cut.
(e) The lower of the two energy deposits was assigned to be the energy deposition from the Compton scatter and the higher of the two energy deposits was assigned to be the energy deposition from the subsequent photoabsorption.

The goal of this event selection and processing was to make a fair comparison of the capabilities of the two instruments as defined by the instrument materials and their physical configuration. The sequencing of the two energy deposits into the first scatter deposit and the second absorption deposit in particular is crude and can result in incorrect sequencing 50% of the time. However, of the simple sequencing options, the one based on assigning the scatter to the lower of the two energies results in the best imaging performance for both of the instruments under comparison. Future work will report on optimization of sequencing for $4\pi$ image production and imager intercomparison. For both instruments, improved imaging would also be possible with imaging algorithms which account for the individual detector responses.

## 2.6. Image reconstruction

After event selection and sequencing, the remaining two-hit events were processed to produce a) the spectrum of the sum of the two energies, $E_{TOT}$, b) the distribution of the Angular Resolution Measure (ARM), where ARM is defined as the difference between the calculated Compton scattering angle and the actual angle between the axis of the two energy depositions and the direction to the source, and c) the Compton back-projection image. Compton back-projection images are shown in the full $4\pi$ of space surrounding the imager. Coloured contours are always shown at the same fixed percentages of the maximum in the image. A threshold is applied such that contours below 50% of the maximum in the back-projection image are not displayed. This 50% threshold requirement has been chosen based on the experience of both teams of developers in producing a product which is least likely to fool non-expert users into believing there is a source in the presence of statistical fluctuations or detector effects in the image.

## 3. RESULTS

### 3.7. Single source results

Fig. 3 shows the result of the common analysis for the data from the H3D H420 detector with the source at -20º off axis. Two minutes of data acquisition are shown here with the H3D 420 detector in the approximately 118 nSv/hr field from the Cs-137 source. There is significant statistical scatter apparent in the Compton back-projection image shown in in Fig. 3 a) yet the image could be easily interpreted by operators in a field situation. The image is consistent with that created by a single radioactive source at a location near the maximum in the back-projection image and this location is consistent with the known true source location. The spectrum of the summed energies of the two hits is shown in Fig. 3 b). The photopeak at 662 keV is extremely sharp and narrow allowing a tight selection for the photopeak events. The angular resolution as indicated by the ARM shown in Fig. 3 b), is quite broad as could be expected from the small size of the sensitive volume of the H3D H420 imager.

The SCoTSS 3×3 result for the same dataset is shown in Fig. 4. The imaging result can be seen in Fig. 4 a) and is much narrow and smoother than the image delivered by the H3D H420 detector under the same trial and processing conditions. The 662 keV photopeak is clearly evident in Fig. 4 b) on which the lines indicating the photopeak selection for image reconstruction are shown. The much higher efficiency of the SCoTSS 3×3 detector, largely due to its greater mass, can be seen in the y-axis values of the $E_{TOT}$ and ARM distributions and this is the major reason for the greater smoothness of the image in Fig. 4 a). However, a big part of the reason that the SCoTSS 3×3 image is superior is the narrowness of the contours of the Compton back-projection image about the actual source position. Much of this effect is attributable to the excellent angular resolution, shown in Fig. 4 c).

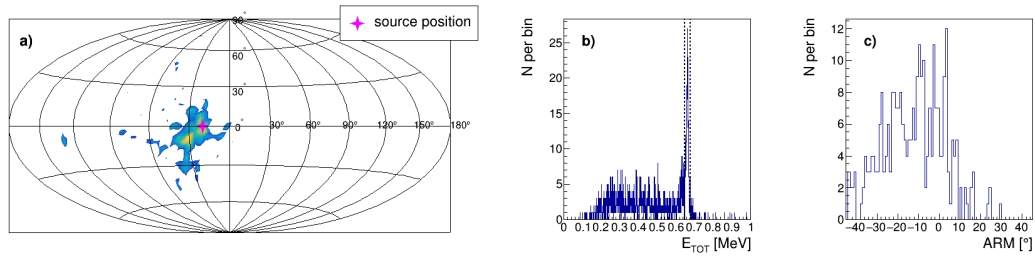

*FIG. 3. H3D H420 result for Run 1, 160 MBq Cs-137 source 10 m away at -20º with two minutes acquisition time. a) 4π display where the coloured contours show the Compton back-projection image. The true source position is indicated by the magenta cross. b) Total energy spectrum for the two-hit events. The vertical dashed lines indicate the photopeak selection window. c) Angular Resolution Measure (ARM) for the two-hit events passing the photopeak selection.*

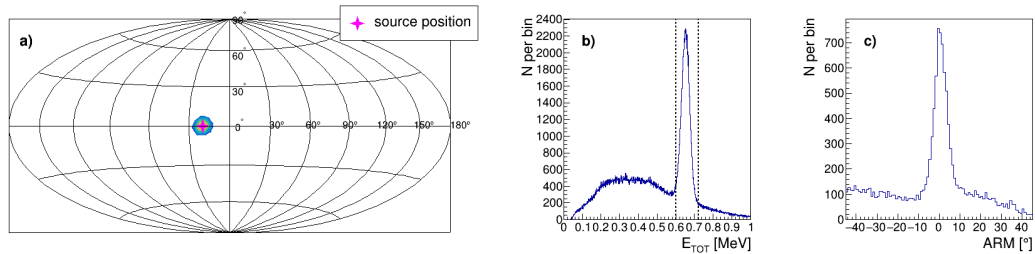

*FIG. 4. SCoTSS 3×3 result for Run 2, 160 MBq Cs-137 source 10 m away at -20º with two minutes acquisition time. a) 4π display showing the Compton back-projection image as the coloured contours. The true source position is indicated by the magenta cross. b) Spectrum of the summed energies of the two hits. The vertical dashed lines show the photopeak selection window. c) Angular Resolution Measure (ARM) for two-hit events passing the photopeak selection and the lever arm requirement.*

The SCoTSS 3×3 imager was designed making use of solid crystalline scintillator such that it could be used in place of the direction-blind NaI(Tl)-based instruments commonly used in aerial and ground-based mobile gamma spectrometry survey. Fig. 5 shows the results using a two-second subset of the data shown in Fig. 4. The information in this result from the perspective of an operator is almost equal to that of the two-minute acquisition period. The Cs-137 photopeak is still clearly evident in Fig. 5 b) and the back-projection Compton image in Fig. 5 a) is tightly centred on the actual source position. Thus, the SCoTSS 3×3 technology is suited for mobile survey and mapping and promising for imaging in motion. The H3D H420 detector on the other hand recorded insufficient events to reconstruct an image with this source in a two-second data acquisition and is therefore not suitable for mobile survey or imaging in motion.

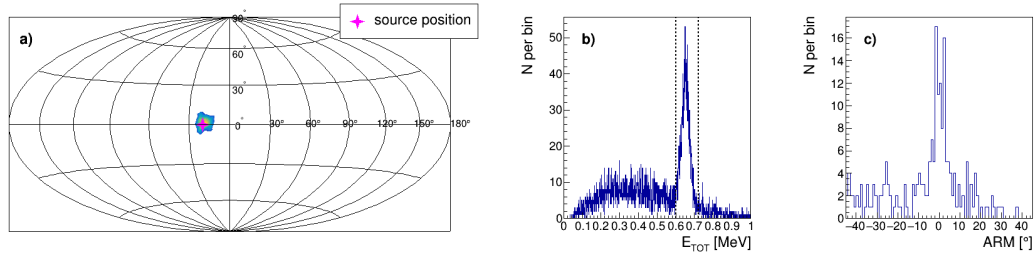

*FIG. 5. SCoTSS 3×3 imager result using 2 seconds of data from Run 2 with a 160 MBq Cs-137 source 10 m away and -20º off axis. a) 4π image. The back-projection Compton image is shown by the coloured contours. The actual source position is indicated by the magenta cross. b) The spectrum of the summed energies of the two hits. The photopeak selection band is indicated by the vertical dashed lines. c) Angular Resolution Measure (ARM) for the two-hit events passing the photopeak and lever arm selections.*

Fig. 6 shows the result from H3D H420 where now the source has been moved to -120º which is to the side of the imager and just behind its nominal forward direction. The H3D H420 has a fairly uniform response throughout 4π space which results in the energy spectrum, angular resolution measure distribution, and 4π image with the source at -120º being of similar quality to that with the source at -20º. We will see that this provides important benefits for imaging of multiple or extended sources.

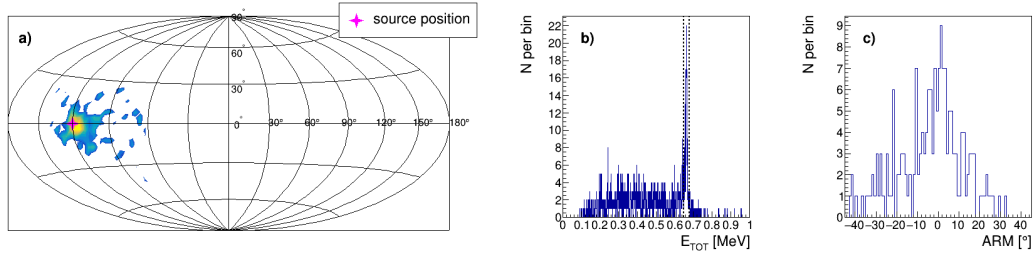

*FIG. 6. H3D H420 result for two minutes of data acquisition in Run 3 with 160 MBq Cs-137 source 10 m distant and -120º off axis. a) 4π image. The Compton back-projection image is shown by the colour contours. The source position is indicated by the magenta cross. b) The spectrum of the sum of the energies of the two hits. The photopeak energy selection is indicated by the vertical dashed lines. c) Angular Resolution Measure (ARM) for the two-hit events passing the photopeak energy selection.*

The SCoTSS 3×3 result for the single Cs-137 source at -120º is shown in Fig. 7. The SCoTSS 3×3 imager, despite being of the two-plane variety originally intended for imaging of sources in the forward direction, successfully localizes the point source which is to the side and rear, as shown in Fig. 7 a). However, note that the efficiency is slightly lower than it was for the source in front, and the angular resolution measure is much broader. This results in broader contours in the back-projection image and we will see that this has detrimental consequences for the reconstruction of multiple or extended sources.

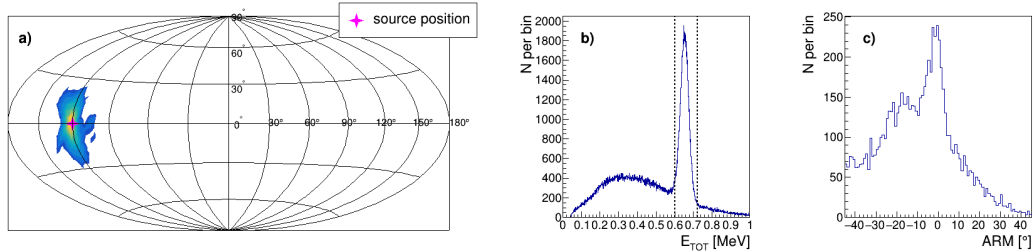

*FIG. 7. SCoTSS 3×3 result for two minutes data acquisition in Run 4 with the 160 MBq Cs-137 source at -120º off-axis, 10 m away. a) 4π Compton back-projection image. The true source position is indicated by the magenta cross. b) Spectrum of the summed energies of the two hits for two-hit events. The vertical dashed lines show the photopeak energy selection. c) Angular Resolution Measure (ARM) for the two-hit events passing the photopeak and lever arm selections.*

### 3.8. Two source results

Fig.'s 8 a) and b) show the 4π imaging results from Runs 5 and 6 with a 10º separation between the two actual source positions. The image from the H3D H420 detector shown in Fig. 8 a) does appear extended as compared to the single source image. However, given the low efficiency and broad angular resolution, it is difficult to be convinced that there is more than one source from this image. The image from the SCoTSS 3×3 detector is shown in Fig. 8 a). Given the high efficiency and angular precision of this device it is possible to observe an elongated shape in the back-projection image which is inconsistent with a single point source. It is not however possible to distinguish by eye between two sources 10º apart, and a source with an elongated shape spanning 10º.

The reconstructed images from an experimental configuration with two Cs-137 sources, one at -20º and one at -50º are shown in Fig. 9. A maximum is reconstructed by the H3D Polaris detector at the -50º location. The image extends across the field of view over several tens of degrees which suggests correctly that the source is widely dispersed rather than a single point source, although the discrimination of two distinct sources of radioactivity is not achieved. The image from the SCoTSS 3×3 detector, shown in Fig. 9 b), on the other hand accurately shows two precise positions for the two sources, one at -20º and one at -50º. The difficulty with the SCoTSS 3×3 detector is that the efficiency and angular resolution vary strongly over the field of view. This causes the source at -50º to appear to be weaker than the source at -20º, while they are in fact nearly the same strength.

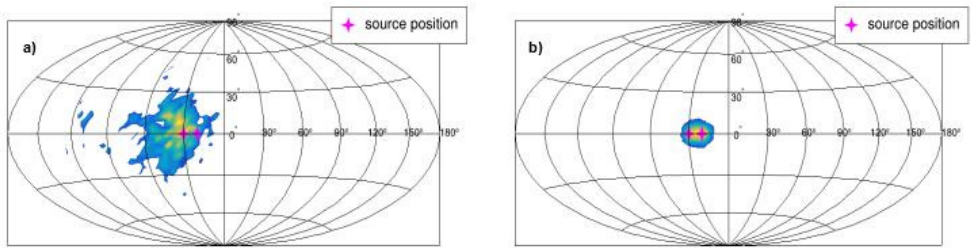

*FIG. 8. 4π images for two minutes of data acquisition in Runs 5 and 6 with two 160 MBq Cs-137 sources 10 m away, one at -20º off axis and one at -30º off axis. The Compton back-projection image is shown by the coloured contours and the actual source positions are indicated by two magenta crosses. a) Run 5 H3D H420 result. b) Run 6 SCoTSS 3×3 result.*

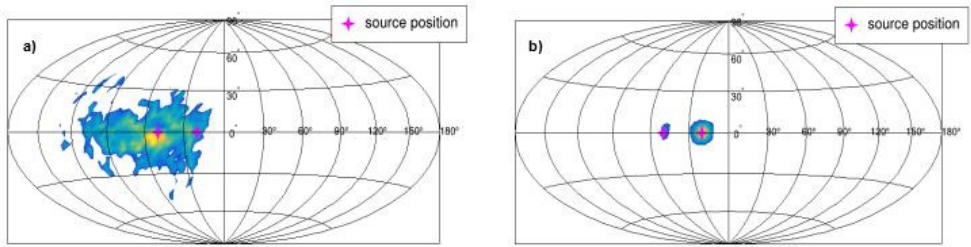

*FIG. 9. Images for Runs 7 and 8 with two sources thirty degrees apart at -20º and -50º. Sources are 10 m away and acquisition time is two minutes. Compton back-projection image is given by the coloured contours. The actual source positions are indicated by the magenta crosses. a) Run 7 H3D H420 result. b) Run 8 SCoTSS 3×3 result.*

Finally, we present images for the two detectors for two sources 100 degrees apart, at -20º and -120º in our coordinate system. The H3D H420 result, shown in Fig. 10 a) successfully indicates the extent of the radioactivity from -120º to -20º, however it is not able to isolate the two sources. The SCoTSS 3×3 image, Fig. 10 b) shows only the source which is at -20º. In fact, the image of the source at -120º must still be there as the source was clearly evident in Fig. 7 a). It is the 50% threshold on display of the colour contours which suppresses the display of the peak at -120º in Fig. 10 b). For the SCoTSS 3×3 technology it should be possible to correct the image for the variation of efficiency and angular resolution across the field of view, to enable the display of the source at 120º, however a robust and simple solution to deal with this problem for any source energy or configuration may be a redesign of the crystal arrangement to one with a uniform response function.

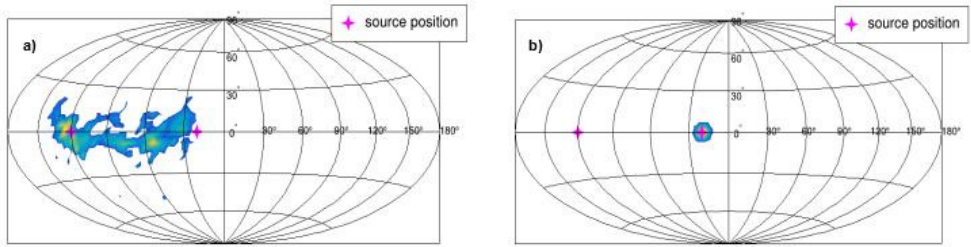

*FIG. 10. Imaging results for Runs 9 and 10, two 160 MBq Cs-137 sources, one at -20º and one at -120º. Sources are both 10 m away and acquisition time is 2 minutes. 4π Compton back-projection images are shown by the coloured contours. The two true source positions are indicated by the magenta crosses. a) Run 9 result with H3D H420 detector. b) Run 10 result with SCoTSS 3×3 detector.*

## 4. DISCUSSION AND CONCLUSIONS

We have presented a limited comparison of gamma imagers based on widely different technologies, semiconductor and scintillator gamma spectrometers. Only two instruments have so far participated in the intercomparison, and trial positions have been limited to source positions on the plane of the horizon of the instruments. The data have not yet been fully analyzed using quantitative measures, and the conclusions herein are based only on the qualitative examination of the distributions and images. Nevertheless, this preliminary work serves to illustrate the value of side-by-side intercomparison in fairly evaluating the strengths and weaknesses of different technological approaches to gamma imaging.

It is necessary for different end users to consider the strengths and weaknesses of different imaging technologies to choose a suitable device for their work. Wide-area mobile survey operations currently employ crystalline scintillator gamma spectrometers of volume from 4 L up to 32 L or more from which a result is required within one second of data accumulation at each grid position of the survey. The SCoTSS technology discussed herein is poised to function as a drop-in replacement for these mobile survey spectrometry systems, bringing the enhancement of greatly improved position reconstruction to the resulting map or image product. In a tiered response, in-situ survey performed by pedestrian teams typically follows the first wide-area mapping. In-situ teams employ high energy-resolution solid-state detectors to study in more detail the isotopic composition of the hot spots identified by the wide-area search. In these follow-on in-situ surveys, the solid-state detector can remain fixed for periods of fifteen minutes or more and replacement of the ubiquitous direction-blind hyper-pure germanium detectors with a semiconductor-based imager such as the H3D H420 could bring the benefits of narrowing the investigation to a certain part of the image, to enhance the ratio of signal to noise even more. All Compton imagers are capable of $4\pi$ imaging and can localize a point source anywhere in space. However, an imager which provides a response function which is flat with source location, should be preferred if the situation to be faced could involve multiple or distributed sources.

## 5. OUTLOOK

Any kind of gamma spectrometer could be adapted to become a Compton imager by incorporating instrumentation to record the locations and amounts of the individual energy deposits. However, Compton imaging does not solve the problem that a certain quantity of material is necessary in order to provide sufficient stopping power. Therefore trade-offs will continue to be made between the size and weight of the spectrometer and the rapidness with which it produces the result and the future will hold as many different kinds of gamma imagers as there are currently gamma spectrometers. As technology advances and the cost and size of the readout and processing electronics used by these instruments decreases, and the ease with which ancillary sensors can be integrated increases, the non-direction-capable instruments will each eventually be replaced by their imaging counterparts. This will bring increased spatial precision and new abilities to localize inaccessible sources for mobile survey, and improved spectrometry capabilities for fixed in-situ survey.

## ACKNOWLEDGEMENTS

The authors gratefully acknowledge the participation of H.Moise and R.Berg from Defence Research and Development Canada's Suffield Research Centre in the trials, particularly with data acquisition using the H3D H420 detector. The authors are grateful to C.Wahl from H3D for general discussions on methods for imager intercomparison. This work was supported in part by Defence Research and Development Canada's Centre for Security Science, project CSSP-2018-TI-2390 and earlier projects. This is NRCan Contribution No. 20190336.